\documentclass{ws-ijmpa}

\begin{document}

\markboth{Henri Bachacou}
{Measurements of the $t \bar t$~production cross section at CDF using b-tagging}

\catchline{}{}{}{}{}
%
%%%%%%%%%%%%%%%%%%%%%%%%%%%%%%%%%%%%%%%%%%%%%%%%%%%%%%%%%%%%%%%%%%%%

\title{Measurements of the $t \bar t$~production cross section at the Tevatron Run~II CDF experiment using b-tagging}

\author{\footnotesize Henri Bachacou}%\\
%on behalf of the CDF collaboration}

\address{Lawrence Berkeley National Laboratory \\
1 cyclotron road, Bldg 50B-5239 \\
Berkeley CA 94720 \\
%U.S.A.
}

\maketitle

\pub{Received (Day Month Year)}{Revised (Day Month Year)}

\begin{abstract}
We present measurements of the $t \bar t$~production cross section in b-tagged lepton + jets events 
from $p \bar p$ collisions at 1.96 TeV using the CDF detector at Fermilab. B-jets are tagged
with either a secondary vertex algorithm, or a soft lepton tagger that identifies muons from B hadron semileptonic decays.
With Tevatron Run II data, we estimate the 
$t \bar t$~signal fraction in two different ways: by estimating the various background contributions, 
and by fitting directly the leading jet transverse energy spectrum for the signal and background contributions. A subset 
of the sample, with two secondary vertex tagged jets, yields a production cross section consistent with the inclusive measurements. 
Results are consistent with a Standard Model $t \bar t$ signal and current measurements of the top quark mass.

\end{abstract}

\section{Introduction}	%) A SECTION HEADING

The Tevatron (Run II) collides protons and anti-protons
head-on at a center-of-mass energy of 1.96 TeV. In such
collisions, the Standard Model (SM) predicts a $t\bar t$ production cross section of
$\sigma_{t \bar t} = 6.7^{+0.7}_{-0.9}$ pb at $m_{\rm t} = 175 \,{\rm GeV}/c^2$~\cite{theory}.
Top quarks are expected to decay almost exclusively to a W boson and a b quark. 
When one W decays leptonically, the $t \bar t$ 
event contains a high transverse momentum lepton, missing energy from 
the unrecorded neutrino, and 4 high transverse momentum jets, 2 of which 
originate from b quarks. We use this decay channel to measure the total $t\bar t$ production cross section.
A deviation from the predicted value would be an indication of new physics either in the production
mechanism or in the top decay. 
We select events with an isolated electron E$_T$ (muon P$_T$) greater than 20 GeV, missing
E$_T>$20 GeV and at least 3 jets with E$_T>$15 GeV and $|\eta| < 2.0$.
Finally, we require at least one jet in the event
to be identified as a heavy flavor jet, either using a secondary vertex algorithm (SECVTX),
or a soft lepton tagger (SLT) that identifies muons from B hadron semileptonic decays.
The analyzes using SECVTX (resp. SLT) are based on $162\,\mathrm{pb}^{-1}$ (resp. $194\,\mathrm{pb}^{-1}$) of data.
The CDF detector is described in detail elsewhere~\cite{CDF}.

\section{\label{sec:secvtx} Measurement with secondary vertex b-tagging.}

%We calculate the cross section as follows:
%\begin{equation}
%\sigma_{t\bar{t}}=\frac{N_{obs}-N_{bckg}}{A_{t\bar{t}}\times
%\int\mathcal{L} dt} \label{eq:xsec}
%\end{equation}
%
%where $N_{obs}$ is the number of events with $\ge 3$ tight jets
%that are tagged with at least 1 SecVtx tag, $N_{bckg}$ is the
%expected background, $A_{t\bar{t}}$ is the total acceptance
%including the b-tagging efficiency, and $\int\mathcal{L} dt$ is the integrated luminosity.

\begin{figure}
\begin{minipage}[t]{0.48\linewidth}
\centerline{\psfig{file=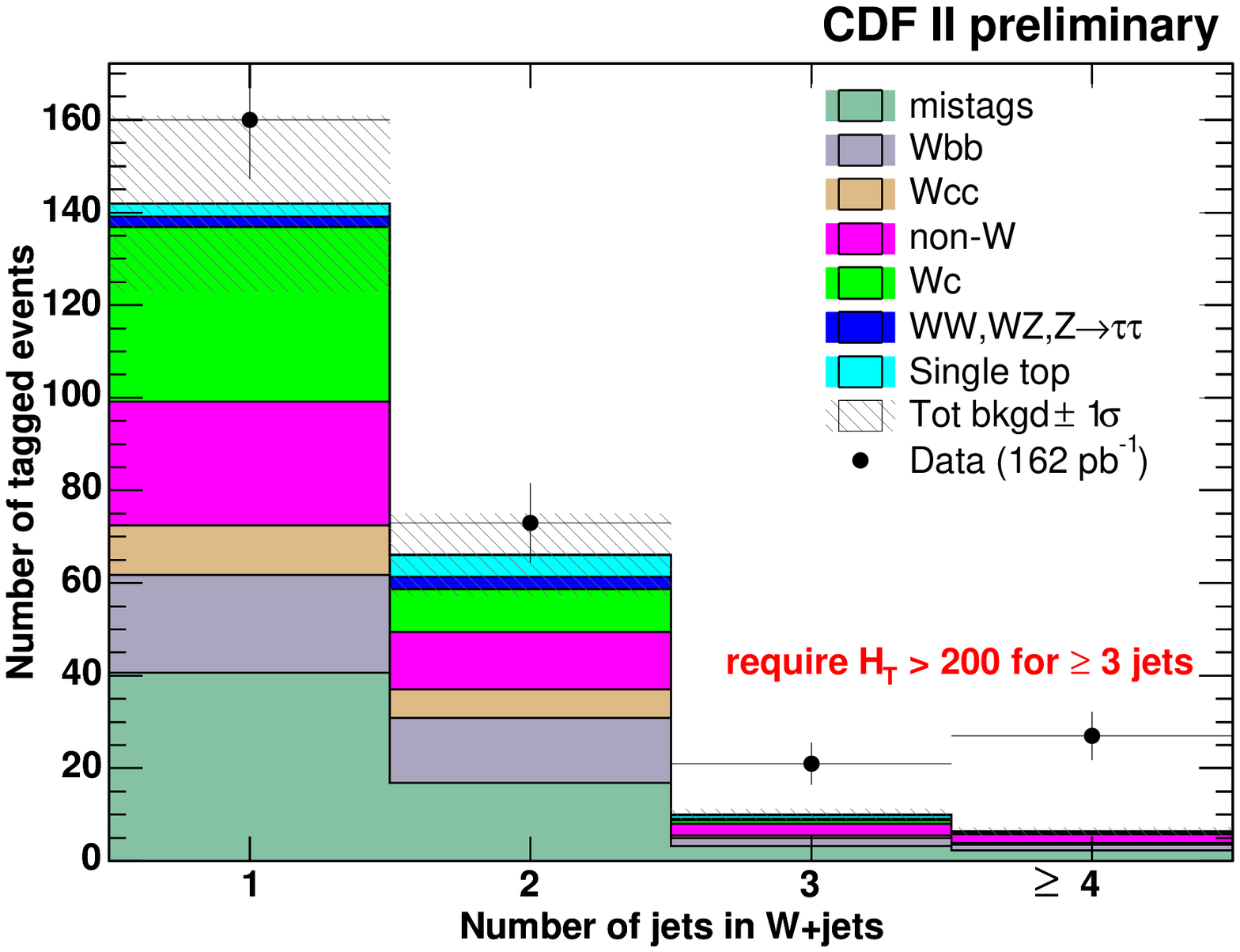,width=6.5cm}}
\vspace*{8pt}
\caption{\label{fig:njets_tags} Jet multiplicity of W+jets events tagged with the SecVtx algorithm
in 162~$pb^{-1}$ of data. The $H_T>200$~GeV cut is only applied to events with three or more jets.}
\end{minipage}
\hspace{.2cm}
\begin{minipage}[t]{0.48\linewidth}
\centerline{\psfig{file=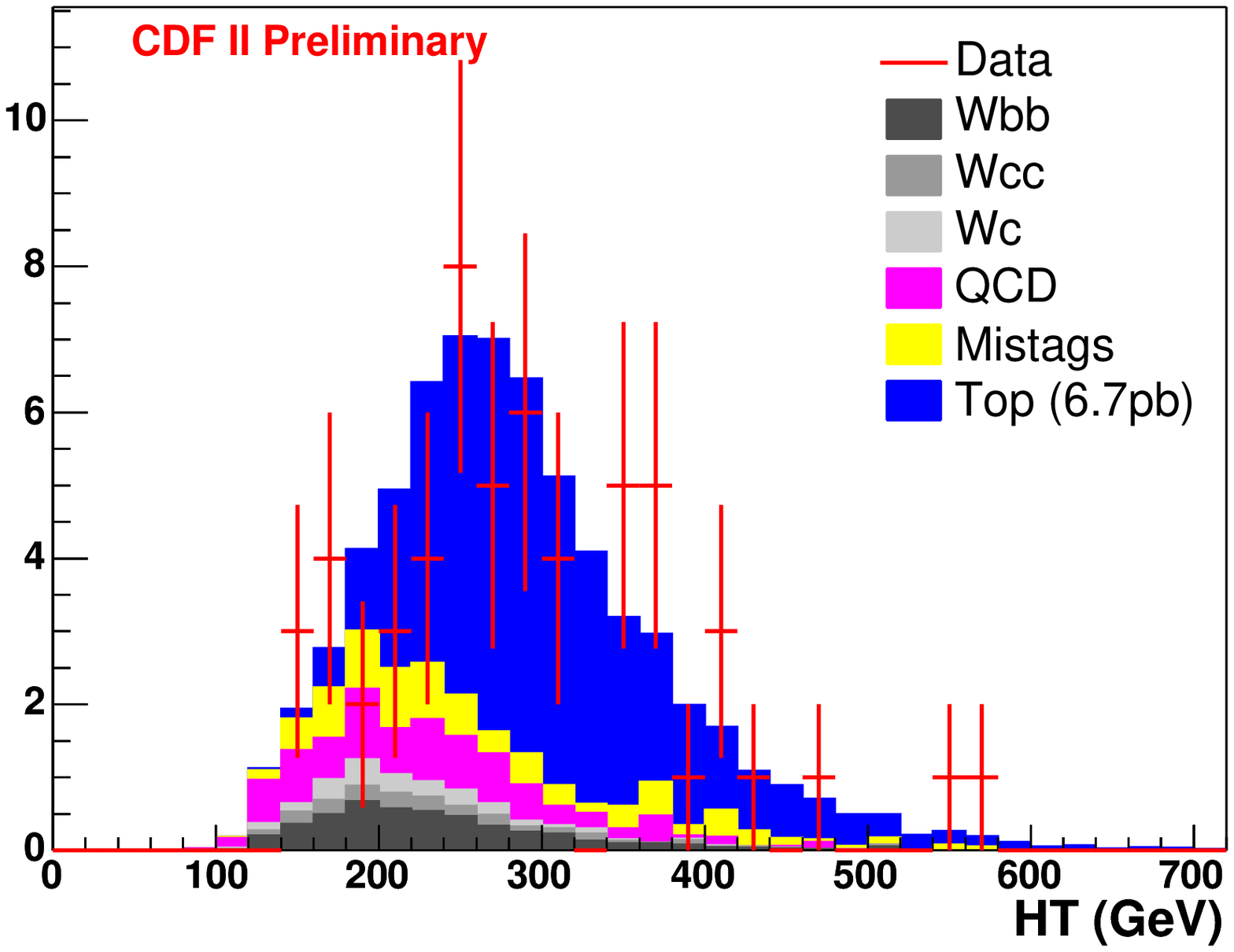,width=6.5cm}}
\vspace*{8pt}
\caption{\label{fig:ht} $H_T$ distribution of the candidate events compared to the expected backgrounds and $t \bar t$ signal.}
\end{minipage}
\end{figure}

We optimize the event selection by requiring that the total transverse energy in the event ($H_T$,
the scalar sum of all jets $E_T$, lepton $p_T$, and missing $E_T$) be larger than 200~GeV.
The SECVTX algorithm selects tracks within the jet
with large impact parameter to reconstruct secondary vertices. Jets containing a secondary vertex 
more than $3\sigma$ away form the primary vertex (in the plane transverse to the beam) are identified as b-jets.
After tuning the simulation on a control sample, the efficiency for tagging
at least one jet in a $t \bar t$ event that passes all other selection requirements
is ($53 \pm 4$)~\%. 
The main sources of background are W + Heavy Flavor events, W + light jets events where
one jet is wrongly tagged, and QCD events that fake a W signal; they are estimated
with techniques that use both Monte Carlo and data control samples.
We expect $13.8\pm2.0$ background events and observe 48 events in the data; 
we measure a cross section of
$5.6^{+1.2}_{-1.1}\rm{(stat.)}^{+1.0}_{-0.7}\rm{(syst.)}\,\rm{pb}$.
Fig.~\ref{fig:njets_tags} shows the number of candidate events vs jet multiplicity
together with the expected background contributions.  
Fig.~\ref{fig:ht} shows the $H_T$ variable distribution of the candidates
compared to the expected background and $t \bar t$~signal (normalized to 6.7 pb).

The sub-sample of events with at least two tagged jets contains 8 events, compared
to an expected background of $0.9\pm0.2$~events, from which we measure a cross section of
$5.4 \pm 2.2 \rm{(stat.)} \pm 1.1 \rm{(syst.)}\,\rm{pb}$.

\section{Measurement with SECVTX using a kinematic fit.}

Instead of explicitly evaluating the contribution to the sample from backgrounds, one can
extract the $t \bar t$ fraction by fitting some kinematic variable in the data. The leading
jet $E_T$ variable was chosen for this purpose. Template shapes for the background
are evaluated from the data; the template shape for $t \bar t$ is from Monte Carlo.
The fit (Fig.~\ref{fig:fit}) measures a $t \bar t$ fraction of $(67^{+13}_{-16})$~\%, leading to a cross section of 
$6.0^{+1.5}_{-1.8}\rm{(stat.)} \pm 0.8 \rm{(syst.)}\,\rm{pb}$.

\section{Measurement with soft muon b-tagging.}

\begin{figure}
\begin{minipage}[t]{0.48\linewidth}
\centerline{\psfig{file=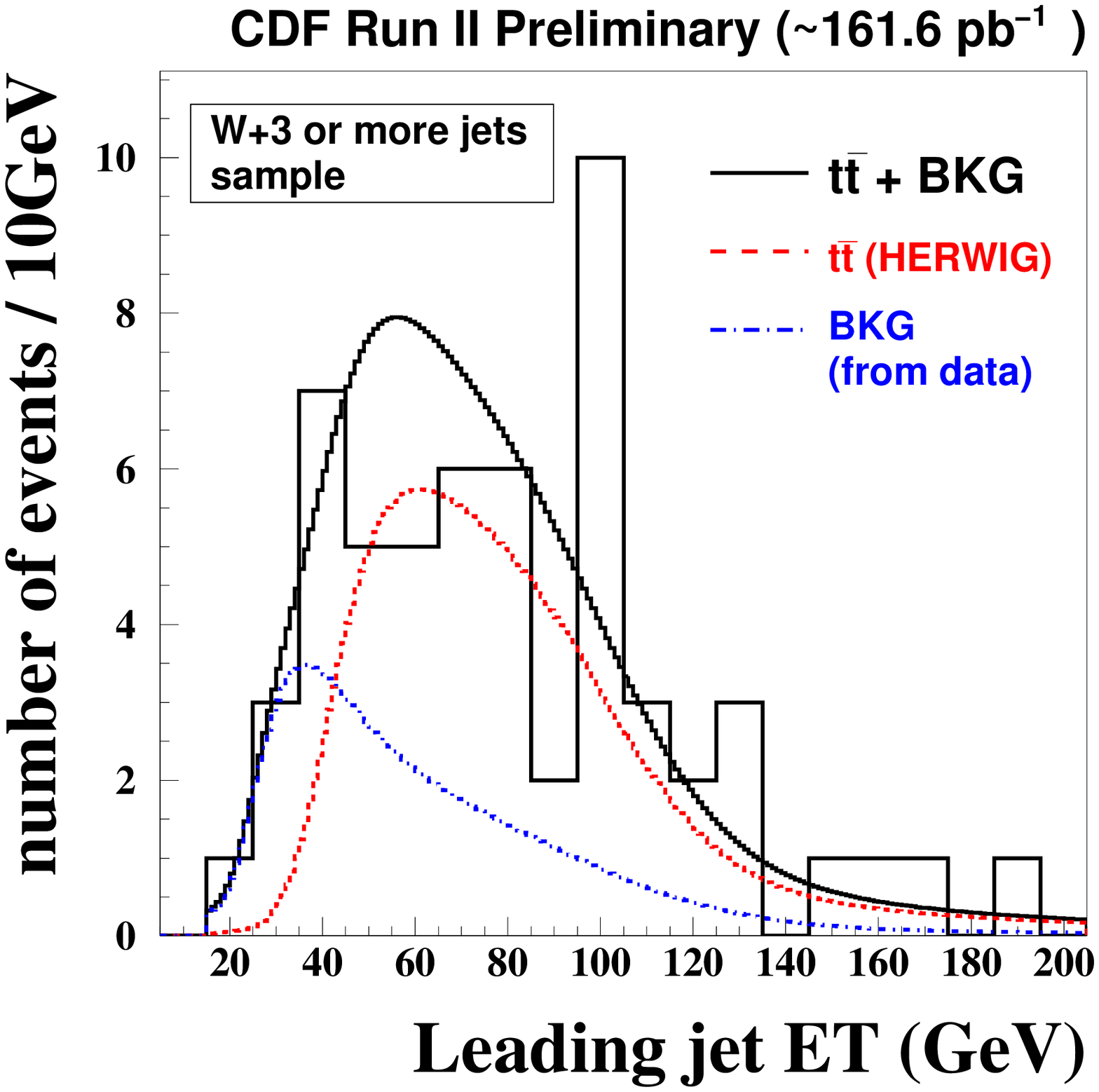,width=6.5cm}}
\vspace*{8pt}
\caption{\label{fig:fit} Leading jet transverse energy of candidates 
in 162~$pb^{-1}$ of data, together with fitted contribution
from $t \bar t$ signal and background.}
\end{minipage}
\hspace{.2cm}
\begin{minipage}[t]{0.48\linewidth}
\centerline{\psfig{file=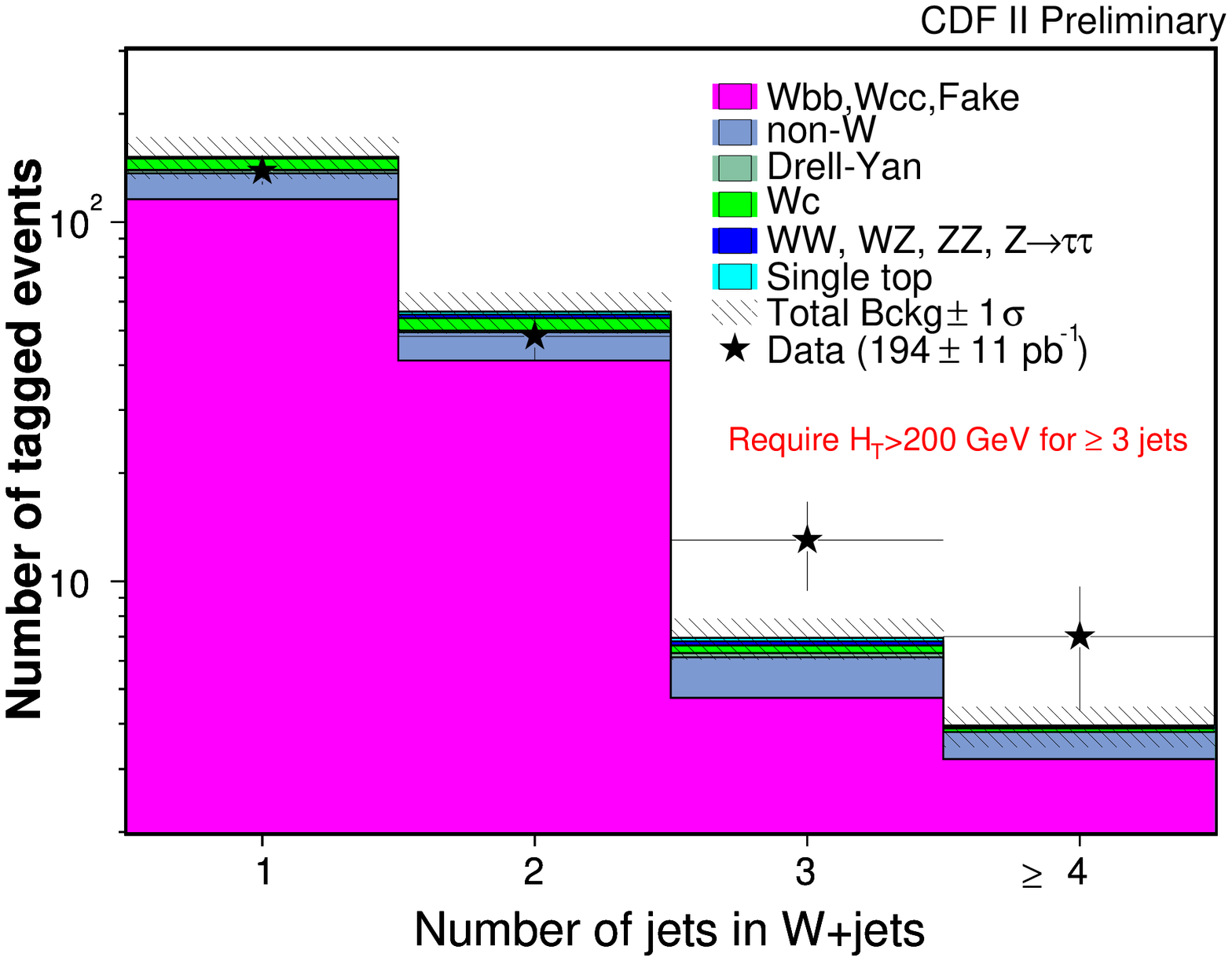,width=6.5cm}}
\vspace*{8pt}
\caption{\label{fig:slt} Jet multiplicity of W+jets events tagged with the SLT algorithm in 162~$pb^{-1}$ of data.}
\end{minipage}
\end{figure}

The muon SLT algorithm matches tracks in the central drift chamber with segments in 
the muon chambers. It uses a global $\chi^2$ built from the matching distributions, to define a
pseudo-likelihood variable, $L$, that separates muon candidates
from background. A jet is considered "tagged" if
it contains an SLT muon with P$_T >$ 3 GeV/c, with $L <$3.5 and
within $\Delta R <$0.6 of the jet axis. Efficiency and fake rate are measured on control samples.
Backgrounds are estimated with techniques similar to Sec.~\ref{sec:secvtx}. 
We expect $11.6\pm1.5$ background events, and observe 20, and we measure a cross section of
$4.2^{+2.9}_{-1.9}\rm{(stat.)} \pm 1.4 \rm{(syst.)}\,\rm{pb}$.
Fig.~\ref{fig:slt} shows the jet multiplicity of the candidates compared to the expected background.

\end{document}